\documentclass[prl,superscriptaddress,showpacs,showkeys,twocolumn]{revtex4-1}
\usepackage{graphicx}				
								
\usepackage{epsfig}
\usepackage{natbib}										
\usepackage{amssymb}
\usepackage{amsthm}
\usepackage{amsmath}
\usepackage{amsfonts}
\usepackage{bm}
\usepackage{color}

\newcommand{\be}{\begin{equation}}
\newcommand{\ee}{\end{equation}}
\newcommand{\bea}{\begin{eqnarray}}
\newcommand{\eea}{\end{eqnarray}}

\def\<{\langle}
\def\>{\rangle}

\usepackage{amsmath,amsthm,amstext,amscd,amssymb,euscript,mathrsfs}
\usepackage{calrsfs}
\usepackage{epsfig}
\usepackage{color}
\usepackage{url}

\renewcommand{\phi}{\varphi}

\def\1{{\mathchoice {\rm 1\mskip-4mu l} {\rm 1\mskip-4mu l}
{\rm 1\mskip-4.5mu l} {\rm 1\mskip-5mu l}}}

\newtheorem*{theorem*}{{\small T}{\scriptsize HEOREM}}
\newtheorem{corollary}{{\bf{\small C}{\scriptsize OROLLARY}}}[section]
\newtheorem*{proposition*}{{\bf{\small P}{\scriptsize ROPOSITION}}}
\newtheorem{lemma}{{\bf{\small L}{\scriptsize EMMA}}}[section]
\newtheorem{remark}{{\bf{\small R}{\scriptsize EMARK}}}[section]
\newtheorem{definition}{{\bf{\small D}{\scriptsize EFINITION}}}[section]

\renewenvironment{proof}[1]
{\noindent{{\bf{\small{ P}{\scriptsize ROOF}}}.}\hspace{0.1cm} #1} {$\;\qed$\newline}

\newcommand{\beq}{\begin{eqnarray}}
\newcommand{\eeq}{\end{eqnarray}}

\newcommand{\ba}{\begin{align*}}
\newcommand{\ea}{\end{align*}}

\newcommand{\bl}{\begin{lemma}}
\newcommand{\el}{\end{lemma}}

\newcommand{\br}{\begin{remark}}
\newcommand{\er}{\end{remark}}

\newcommand{\bt}{\begin{theorem}}
\newcommand{\et}{\end{theorem}}

\newcommand{\bd}{\begin{definition}}
\newcommand{\ed}{\end{definition}}

\newcommand{\bc}{\begin{corollary}}
\newcommand{\ec}{\end{corollary}}

\newcommand{\bpr}{\begin{proof}}
\newcommand{\epr}{\end{proof}}

\newcommand{\bi}{\begin{itemize}}
\newcommand{\ei}{\end{itemize}}

\newcommand{\ben}{\begin{enumerate}}
\newcommand{\een}{\end{enumerate}}

\newcommand{\cg}[1]{\textcolor[rgb]{0,0,0}{#1}}

\begin{document}

\title{Non-equilibrium 2D Ising model with stationary uphill diffusion}

\author{Matteo Colangeli}
\affiliation{University of L'Aquila, Via Vetoio, 67100 L'Aquila, Italy.}

\author{Cristian Giardin\`a}
\affiliation{University of Modena and Reggio Emilia, Via Universit\`a 4, 41121 Modena, Italy}

\author{Claudio Giberti}
\affiliation{University of Modena and Reggio Emilia, Via Universit\`a 4, 41121 Modena, Italy}

\author{Cecilia Vernia}
\affiliation{University of Modena and Reggio Emilia, Via Universit\`a 4, 41121 Modena, Italy}

\date{\today}

\begin{abstract}
Usually, in a non-equilibrium setting, a current brings mass from the highest density regions to the lowest density ones.
Although rare, the opposite phenomenon (known as ``uphill diffusion'') has also been observed in multicomponent
systems, where it appears as an artificial effect of the interaction among components. We show here that
uphill diffusion can be a substantial effect, i.e. it may occur even in single component systems 
\cg{as a consequence of some external work}. 
\cg{To this aim we consider the 2D ferromagnetic Ising model in contact with two reservoirs that fix, at the left and the right boundaries,
magnetizations of the same magnitude but of opposite signs.}
\cg{We provide 
numerical evidence that} a class of non-equilibrium steady states exists in which,
by tuning the reservoir magnetizations, the current in the system changes from ``downhill'' to ``uphill''.
Moreover, we also show that, in such non-equilibrium set-up, the current vanishes precisely when the reservoir 
magnetizations equal the magnetization of the corresponding equilibrium dynamics, thus
establishing a novel relation between equilibrium and non-equilibrium properties.
\end{abstract}



\maketitle 

\noindent
{\bf Introduction.}
When a metal bar is put in contact at its extremity with two heat sources at different temperatures,
heat is transported from one side to the other. Fourier's law \cite{fourier} of heat conduction, $J = - \kappa \nabla T$,  
states that the heat current $J$ is proportional to the temperature gradient  $\nabla T$  and the constant of proportionality 
$\kappa$ defines the thermal conductivity. Fourier's law also provides a {\em minus} sign for the current, whose direction is 
against the temperature gradient (i.e. the heat current flows from the hottest to the coldest side). One then says that the current goes ``downhill''.

Surprisingly, the phenomenon of ``{\em uphill diffusion}'' -- namely a current which goes up the gradient, 
and thus has the {\em ``wrong'' sign} -- 
has been observed in several instances, including experiments measuring the diffusion of carbon in 
austenite metals \cite{larken}, multicomponent mixtures \cite{krishna}, microscopic systems with 
multiple conservation laws \cite{bernardin,olla}. 
The work described here is motivated by 
such unexpected behavior that
seems to contradict the empirical  laws of transport (e.g. Fourier's law for
heat transport or Fick's law for mass transport) whose general validity is based
on the physical property that diffusion is a phenomenon smoothening concentration
gradients. However, in all the previous examples the diffusion flux of any species (or conserved quantity) 
is strongly coupled to that of its partner species. If one focuses 
on one particular species, one sees the other species acting as an
effective external field. \cg{As a result of
this coupling 
uphill transport may occur in one
particular component \cite{Krishna1,Krishna2,Krishna3}}.

In this letter we shall show that {uphill diffusion} may arise as a {\em substantial effect}
in single component systems in the presence of {a phase transition}. 
\cg{In our setting the current flowing in the wrong direction is a consequence
of the work that is performed by external reservoirs.} We shall consider  simplified mathematical models of interacting particle systems
(stochastic lattice gases) in a non-equilibrium stationary state due to a {\em boundary driven current}.
We shall show that in such systems there is uphill diffusion, i.e. the current brings mass from the 
region with the smallest density phase to the one with the largest density.
Some theoretical evidence of this intriguing physical phenomenon was recently reported in 
\cite{anna,CDMPplA2016,CDMPjsp2017,CDMP2017} for 1D particle systems with Kac 
potentials (where phase transitions are obtained in a mean-field limit).
We shall study here the simplest mathematical model of a physical
system displaying a true phase transition, i.e. the 2D Ising model
in a non-equilibrium stationary state. 
To our knowledge, this is the first example of a model 
with a phase transition exhibiting non-equilibrium steady states with uphill diffusion. 

\bigskip

\noindent
{\bf The model and the main result.} 
We consider the non-equilibrium dynamics of the nearest-neighbor ferromagnetic Ising model on a finite squared lattice
$\Lambda$ of linear size $L$ coupled to magnetization reservoirs on the horizontal direction. 
To each lattice site $i\in\Lambda$ we associate a spin variable $\sigma_{i}(t) \in \{-1,+1\}$ that 
describes the microscopic state at time $t$. 
The Ising model is equivalent to a lattice gas model via the standard
mapping between spin variables $\sigma_{i}$ and occupation variables $\eta_i\in \{0,1\}$
($\eta_i = (1+ \sigma_i)/2$) with $\eta_i = 1$ (resp.  $\eta_i = 0$) denoting 
the presence (resp. absence) of a particle.
The spins interact with their nearest neighbors according to the Hamiltonian 
\be
\label{H-Ising}
H(\sigma) = - \frac{1}{2}\sum_{\underset{|i-j| = 1}{i,j\in \Lambda}} \sigma_i \sigma_j,
\ee
\cg{where the boundary conditions are specified below.}
In the infinite volume limit it is well known that
the 2D Ising model has a phase transition at the 
inverse critical temperature
computed by Onsager \cite{onsager}
$$
\beta_c = \frac{\ln(1+\sqrt{2})}{2} \approx 0.440686
$$
For inverse temperatures $\beta>\beta_c$ the model exhibits a spontaneous magnetization
given by the formula \cite{yang}
\be
\label{mbeta}
m_\beta = \left[1-\sinh^{-4}\left(2\beta\right)\right]^{1/8}
\ee

\begin{figure}[h!]
\includegraphics[width=8cm]{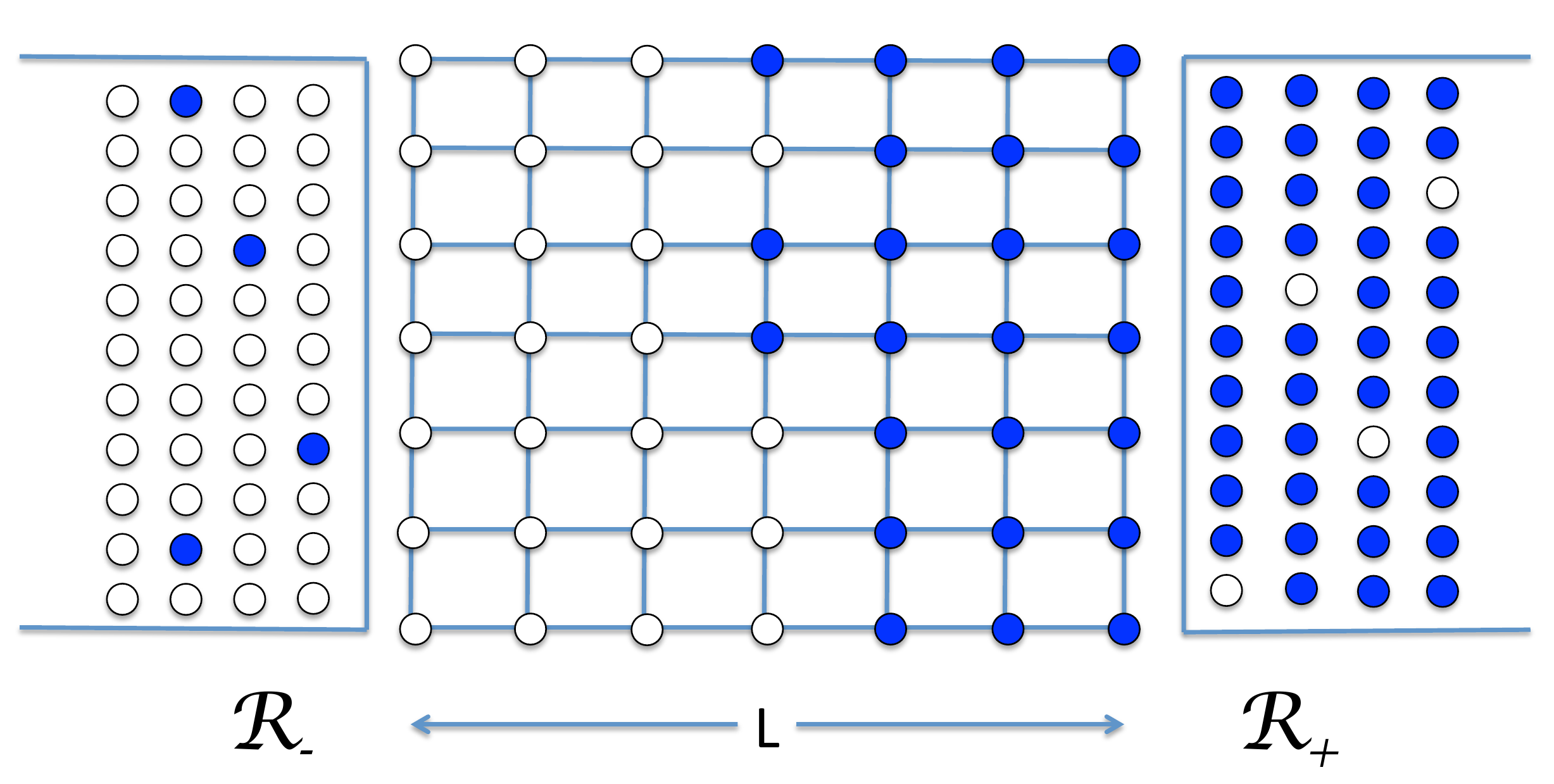}
\caption{Schematic picture of the 2D Ising model coupled to reservoirs 
$\mathcal R_+$ and $\mathcal R_{-}$.
A spins up is represented with a filled (blue) circle, a spin down is represented with
an empty (white) circle.
}
\label{fig0}
\end{figure}

We consider the system in the low temperature region $\beta>\beta_c$ and let the spins evolve 
following a continuos-time stochastic dynamics with two contributions: a conservative exchange 
dynamics in the bulk and independent spin flips at the boundaries. The dynamics at the boundaries 
simulates two infinite reservoirs, $\mathcal R_+$ on the right and $\mathcal R_{-}$ on the left, 
that force a magnetization $ m_+ \in [0,1]$ on the right column and a magnetization $m_-=-m_+$ on 
the left column. 
See Fig. \ref{fig0} for a description of the set-up in numerical experiments.

More precisely,  in the {\em bulk} the spins follow a Kawasaki dynamics, 
i.e. the spins of a bond $\langle i,j\rangle$ exchange values at rate
\begin{equation*}
c(i,j) = \left\{
\begin{array}{ll}
1 & \text{if } \Delta H = H(\sigma^{ij}) - H(\sigma) \le 0\\
e^{-\beta \Delta H} & \text{otherwise}
\end{array} \right.
\end{equation*}
where 
$\sigma^{ij}$ denotes the configuration obtained from $\sigma$ by
exchanging the spins at sites $i$ and $j$.
At the horizontal {\em boundaries} the spins flip independently,
i.e. they change sign at rate
\begin{equation*}
c_{-}(i) = 
\frac{1-\sigma_{i}m_{-}}{2}  \qquad \text{if} \quad i = (1,y)
\end{equation*}
\begin{equation*}
c_{+}(i) = 
\frac{1-\sigma_{i}m_{+}}{2} \qquad \text{if} \quad i = (L,y)
\end{equation*}
Due to the presence of the reservoirs the dynamics  
is not reversible w.r.t. the Boltzmann-Gibbs measure with Hamiltonian \eqref{H-Ising}.
A non-equilibrium steady state sets in characterized by
a uniform current in the horizontal direction. 
A similar setting has been considered in \cite{spohn},
where  the stable region with normal mass transport 
was considered and the fluctuations 
of the interface separating the two phases
were studied. Thus, the focus  in \cite{spohn} was different than in our paper.

As a result of the simulations we observe the following phenomenology:
 as   $m_+$ decreases from $m_+=1$ the current is first {\em negative} and, 
 past a critical value $m_{crit}$, it becomes {\em positive}. We conclude from the  simulations that:
\begin{itemize}
\item If $m_+ > m_{crit}$  then  the magnetization flows from the plus to the minus phase (from $\mathcal R_+$  to $\mathcal R_-$) so that the current is negative (in agreement with the Fick's law) 
and the current goes {\em{downhill}}.
\item If  $m_+ < m_{crit}$ the magnetization flows from the minus to the plus phases (from $\mathcal R_-$  to $\mathcal R_+$), thus the current is positive and 
we have ``{\em uphill diffusion}''.

\end{itemize}
As we shall see, the  value of the critical magnetization 
marking the transition from down- to up-hill diffusion 
$m_{crit}=m_{crit}(\beta,L)$ is a function of both  
the inverse temperature $\beta$ and the system size $L$.
For simplicity, we avoid in the following to write explicitly such dependences.
Our results suggest that in the limit of large boxes
$L\to\infty$ 
the critical magnetization approaches the equilibrium spontaneous magnetization $m_{\beta}$. 

\bigskip

\noindent
{\bf Numerical analysis of the current.}
The integrated current $J_t$ over any horizontal bond up to time $t$ 
can be measured by counting the number of positive spins 
that cross the bond \cg{from left to right} minus the number 
of positive spins that cross the bond in the opposite
direction. The current $J$ in the stationary state is
then obtained as $J = \lim_{t\to\infty} J_t/t$.
We have fixed $\beta=1$ and run computer simulations with $L \le 40$
for various values of $m_+$ and $m_-=-m_+$.
We imposed periodic b.c. on the direction orthogonal to the current. 
Namely, denoting by $i=(x,y)$ the coordinates of site $i$, we set 
$\sigma_{(x,L+1)} = \sigma_{(x,1)}$ for all $x=1,\ldots, L$.
On the longitudinal direction we considered two types of boundary 
conditions: (a) {\em fixed b.c.},  i.e.
$\sigma_{(0,y)} = -1$, $\sigma_{(L+1,y)} = +1$ for all $y=1,\ldots, L$;
(b) {\em shifted b.c.}, namely we let $\sigma_{(1,y)}$ interact with $\sigma_{(1,y-L/4)}$
and  $\sigma_{(L,y)}$ interact with $\sigma_{(L,y-L/4)}$. 
We will explain later this choice of b.c.
(that is inspired by \cite{bodineaupresutti}).
No difference 
in the results obtained using the two different boundary conditions
on the longitudinal direction was observed in our simulations.

We run two independent programs by implementing both the classical
Metropolis Monte Carlo method as well as the kinetic Monte Carlo
method \cite{kratzer}. Whereas the two dynamics \cg{yield} the 
the same 
\cg{stationary state}, the first algorithm is better suited 
to measure the current and the second, which implements a continuous 
time dynamics, is more efficient to probe the magnetization time average.

\begin{figure}[h!]
\centerline{\psfig{file=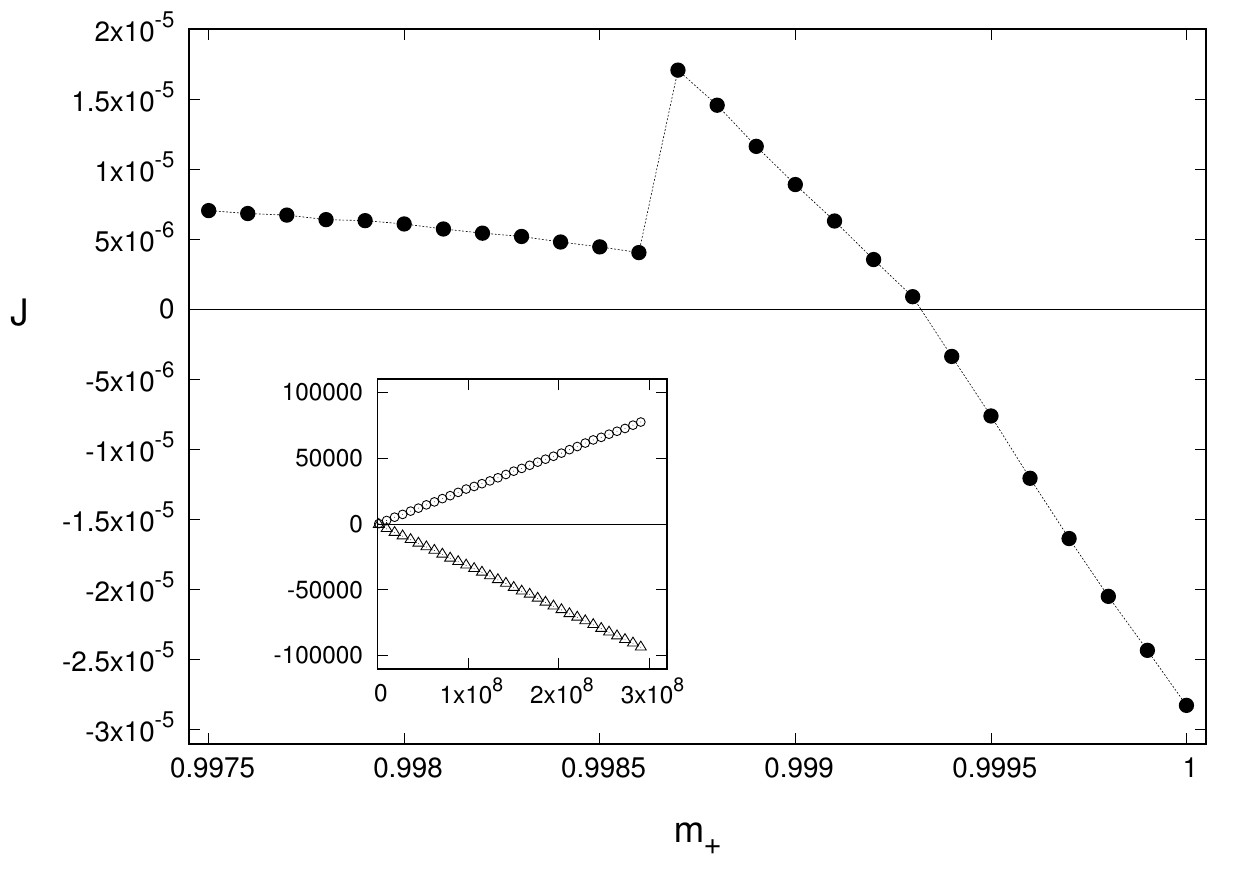,width=8cm}}
\caption{Current vs. reservoir magnetization for system size $L=40$.
Each data point is the current $J$ measured in the non-equilibrium
stationary state with a given value $m_+$ on the right reservoir  $\mathcal R_+$
and $m_- = -m_+$ on the left reservoir $\mathcal R_-$. The inset
shows the integrated current $J_t$ up to time $t = 3 \times10^8$ steps 
for $m_+=0.9995$ (negative slope) and for $m+=0.99910$ 
(positive slope). The initial datum used in the Monte Carlo simulations is: $\sigma_{(x,y)}=-1$ for $x\in[1,L/2]$ 
and $\sigma_{(x,y)}=1$ for $x\in(L/2,L]$.}
\label{figura1}
\end{figure}

Our main result is illustrated in Fig. \ref{figura1}. There it is plotted the 
current $J$ as a function of the right reservoir magnetization $m_+$,
which varies in the interval $[0.9975, 1]$ in steps of $10^{-4}$.
The current has to be measured over a sufficiently
long time span to get rid of fluctuations and to ensure the
convergence to the stationary regime. This can be tested
by monitoring the running average of the current and looking at the
scale of its fluctuations. As a result, we have verified
that $10^{12}$ \cg{spin exchanges} are needed to guarantee fluctuations 
of order \cg{$10^{-7}$} in the worst cases.
\cg{In Fig.  \ref{figura1} errors bars are smaller than the size of the points.}
From Fig.  \ref{figura1} we see the existence of a critical value $m_{crit} \approx 0.99931$
such that if $m_+ > m_{crit}$ then the current is negative,
and if $m_+ < m_{crit}$ the current is positive.
To let better appreciate the change of sign we plot in the
inset the integrated current $J_t$ up to time $t = 3 \times10^8$ steps.  
We see that for $m_+=0.99950$ there is a straight line with a
negative slope, whereas for $m+=0.99910$ we measure a 
positive slope.

In order to gain some understanding on the transition from down- to up-hill diffusion
we start from equilibrium (i.e. the setting without reservoirs) considering the canonical 
Gibbs measure with Hamiltonian \eqref{H-Ising}, inverse temperature $\beta>\beta_c$ 
and total magnetization $m = 0$. 
This is the Wulff problem first studied in \cite{dks}. For a system of large linear size $L$
it is proved in \cite{dks} that the typical configurations have the following structure: there 
is a vertical strip centered at $L/2$ of macroscopically infinitesimal thickness: to the right 
of the strip the magnetization is essentially $m_\beta$ and to the left $-m_{\beta}$ (or 
viceversa). 

In the non-equilibrium setting the interface separating the plus and minus phase
is perturbed by the current originated by the reservoirs, while the optimal magnetization profile
must also interpolate between the value at the right side $m_+$ and its negative
value $m_- = - m_+$ at the left side. When $m_+=1$ one expects that the {\em instanton}
is stable: the magnetization profile $m(r)$ in the macroscopic coordinate \cg{$r=x/L$ (thus $r\in[0,1]$)}  
starts from $m(0) = -1$, for $r<1/2$ increases monotonically to $-m_{\beta}$, 
at $r=1/2$ it has a jump of magnitude $2m_{\beta}$ and finally increases monotonically 
again for $r>1/2$ from  $ m_{\beta}$ to  $m(1) = 1$.
Such profile sustains a negative current, which is microscopically due to positive spins
(resp. negative) that cross the interface from the right (resp. left) and are eventually 
absorbed by the left (resp. right) reservoir.

When $m_+ <1$ a second microscopic mechanism produces a current: 
positive spins (resp. negative) that are created at the left (resp. right) reservoir
and travel to the right (resp. left), thus yielding a positive contribution to the current.
Indeed, we see in Fig. \ref{figura1} that the current increases as $m_+$
is decreased from $1$. At $m_+ = m_{crit}$ the two contributions
to the current of microscopic origin balance themselves, 
thus yielding zero current. Past $m_{crit}$ the positive contribution to
the current is dominant.
\begin{figure}[h!]
\centerline{\psfig{file=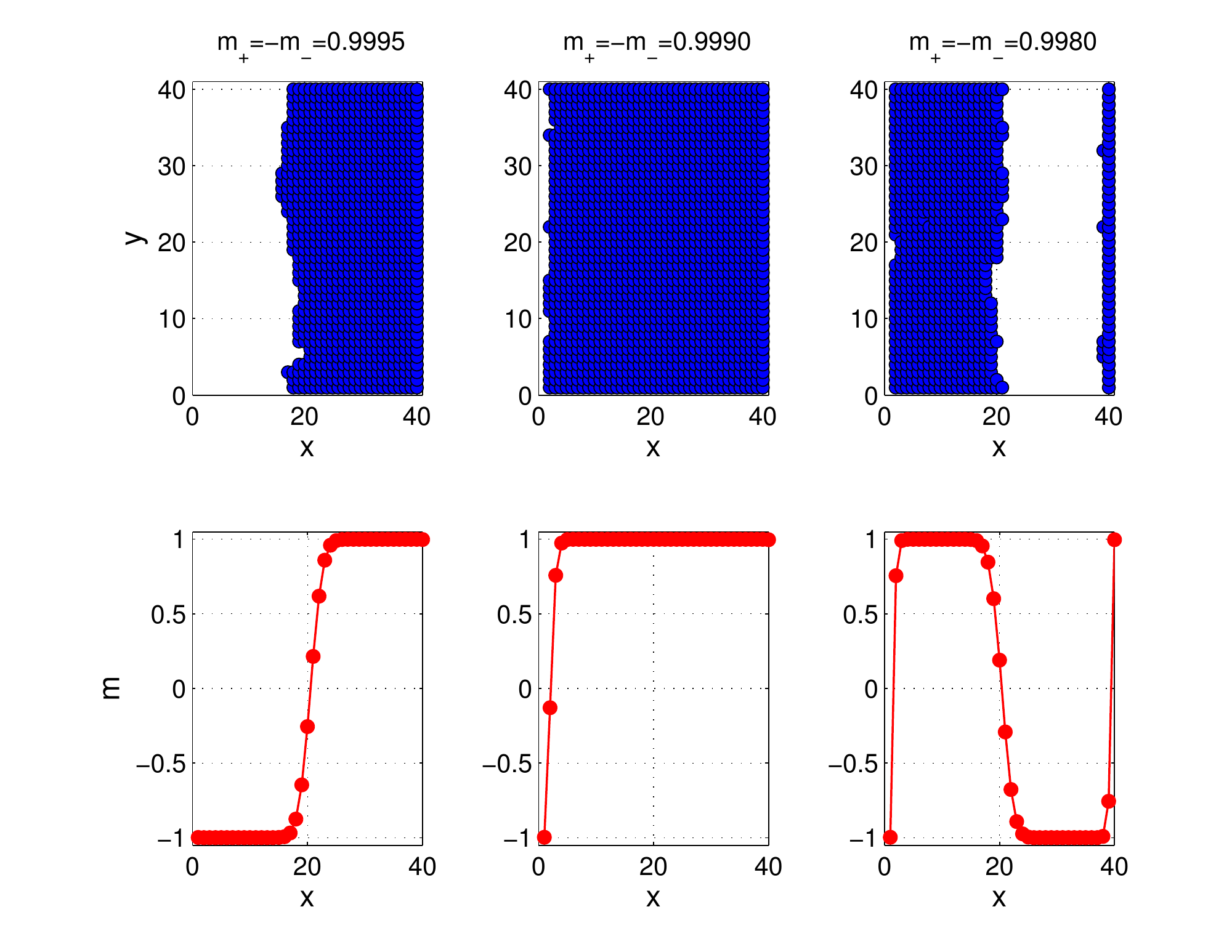,width=8cm}}
\caption{Spin configurations (top panels) and 
time-averaged magnetization profiles (bottom panels) for 
three values of the reservoir magnetization:
$m_+= -m_- = 0.9995$ stable phase (first column);
$m_+= -m_- = 0.9990$ meta-stable phase (second column);
$m_+= -m_- = 0.9980$ weakly-unstable phase (third column). }
\label{figura4}
\end{figure}

The analysis of the typical spin configurations and time-averaged 
magnetization profiles show that past $m_{crit}$ there is a change 
in the structure of the non-equilibrium steady state. 
We run a simulation with kinetic Monte Carlo method doing $10^{10}$ spin
exchanges and plot in Fig. \ref{figura4} the spin configuration at the
end of the run (top panels) and the time averaged magnetization profiles (bottom panels). 
Whereas for $m_+ > m_{crit}$
the non-equilibrium stationary 
\cg{state} is still concentrated on the 
instanton profile (Fig. \ref{figura4}, first column), for $m_+ < m_{crit}$  
we see from the numerical simulations that
the instanton becomes unstable.
Two regimes can be clearly detected:
a metastable phase where the instanton is replaced by a {\em bump} (Fig. \ref{figura4}, second column)
and, continuing to lowering $m_+$,  a weakly-unstable phase appears 
with a profile {\em with two bumps} (Fig. \ref{figura4}, third column).
Note that in Fig. \ref{figura1} the current has a discontinuity around $m_+\simeq 0.9987$, which signals the onset of a \textit{dynamical transition} 
from the ``bump`` typical configuration (in the metastable region) to the ``two-bumps`` configuration (in the weakly-unstable region).
Remarkably, a similar scenario was also observed in \cite[Fig. 14]{CDMPjsp2017} in the case of a 1D particle system equipped with an attractive long-range Kac potential.

\bigskip

\begin{figure}[h!]
\centerline{\psfig{file=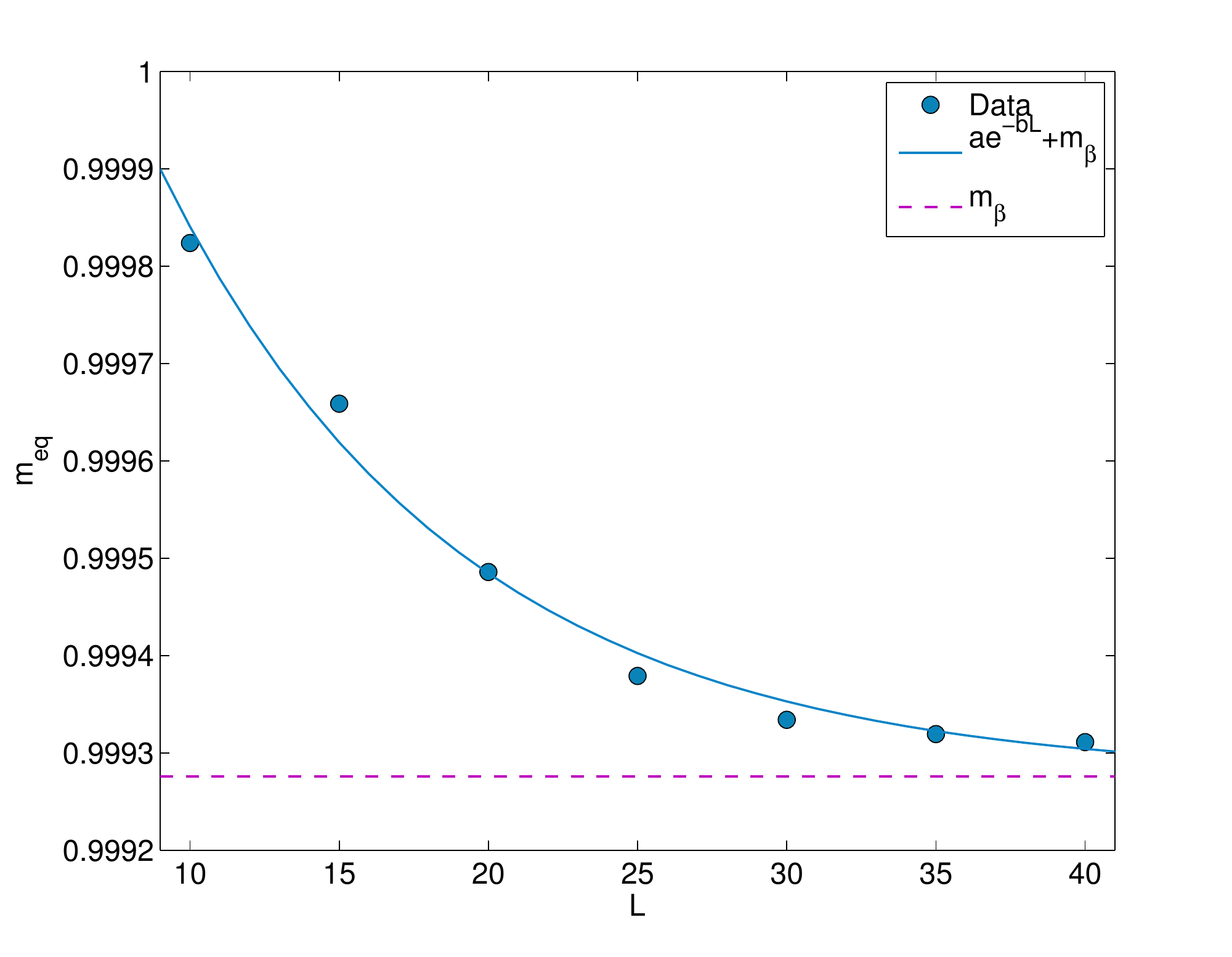,width=8cm}}
\caption{Time-averaged magnetization on the last column $m_{eq}$
at equilibrium (i.e. without reservoirs) versus system size $L$. The horizontal line
is $m_\beta \approx 0.99927$. The continuous line is an exponential fit
$m_{\beta} + ae^{-bL}$ with $a = 0.00153$ and  $b = 0.0996$.}
\label{figura5}
\end{figure}

\noindent
{\bf Estimate of the critical magnetization}.
We claim that the critical value $m_{crit}$ of $m_+$ can be
estimated with an independent method. 
Following the theory given in \cite{bodineaupresutti}, the key quantity is the magnetization value $m_{eq}$ 
{\em on the rightmost column} of the lattice measured at equilibrium, i.e.
in the absence of reservoirs. We claim that  $m_{eq}$ 
must be very close to  $m_{crit}$. Indeed if $m_+ = m_{eq}$ 
(and $m_{-} = -m_{eq}$) then in the non-equilibrium setting the reservoirs 
are trying to impose a magnetization which is already there, so that their
influence is negligible. Therefore the current in the presence of the reservoirs is
essentially the current without reservoirs, which is zero.
The choice of the shifted b.c. guarantees that even close
to the boundaries one would see in a
very large system a magnetization $m_\beta$ to the right of
the interface and $-m_\beta$ to the left.
However when $L$ is finite the magnetization at the boundaries
is not exactly equal to $m_\beta$ due to finite-size effects.
Thus $m_{eq}$ at finite volume might well be different
from $m_\beta$.
For a system size $L=40$ the simulation at equilibrium yields a 
value for $m_{eq} \approx 0.99931$, 
thus in perfect agreement with the value of $m_{crit}$ obtained from
the non-equilibrium simulations. 
We measured the value of $m_{eq}$ for several system 
sizes with $L$ in the range $[10,40]$. We found that 
these values decrease with increasing $L$.
A plot against $L$ is shown in Fig \ref{figura5}, together
with an exponential fit. The extrapolation to the infinite volume is
compatible with an asymptotic value of $m_{eq}$ 
 equal to $0.99927$, that coincides approximatively
with $m_{\beta}$ in \eqref{mbeta} evaluated at $\beta=1$.

\bigskip

\noindent
{\bf Discussion.}
In this paper it is argued that uphill diffusion appears in the
non-equilibrium Ising model coupled to magnetization reservoirs.
\cg{A few final comments are in order.
First we observe that our results imply no violation of the thermodynamic principles.
Indeed, our system (composed of a channel and left/right reservoirs) is not an isolated system.
On the contrary, the Glauber dynamics at the boundaries is such that energy is systematically
pumped into the channel. 
A second issue is the extrapolation to the thermodynamic limit $L\to \infty$.
While this remains an admittedly open issue, we observe that our simulations
with $L=40$ provides perfect agreement between the critical magnetization value 
$m_{crit}$ signaling the onset of uphill diffusion and the magnetization value $m_{eq}$ 
measured at equilibrium on the rightmost column of the lattice.
Furthermore, in the range $L\in[10,40]$ we could verify the expected exponential convergence
of $m_{eq}$ to equilibrium spontaneous magnetization $m_{\beta}$.
All this is evidence that our non-equilibrium simulations are
capable of reproducing the infinite volume equilibrium state
including its finite volume corrections. 
We are currently investing larger sizes \cite{future} to verify the conjecture that
uphill diffusion persists in the thermodynamical limit $L\to \infty$.}

The apparent contradiction between uphill diffusion and the validity
of Fick's law can be resolved by looking at the magnetization profiles. 
Specifically, in the second column of Fig. \ref{figura4} 
we measured a value of magnetization at the peak of the bump
that is between $m_+$ and  $m_{\beta}$, 
namely 
$m_+ = 0.9990 < m_{bump} = 0.99925 < m_{\beta} \approx 0.99927$.
In between the peak and the right boundary, the magnetization
profile is monotonically decreasing,
thus most of the magnetization profile is compatible
with a {\em positive} current that is  down the gradient. 
In the third column of Fig. \ref{figura4} we found instead
$m_+ = 0.9980 <  m_{\beta} \approx 0.99927 < m_{bump} = 0.99940$.
Thus, being $m_{\beta} < m_{bump}$, we have again
downhill current.

It is natural to ask what is the structure of the non-equilibrium stationary
\cg{state} as one continues to lower $m_+$.
We \cg{see} from the simulations that
the weakly-unstable region with a double bump 
persists until, approximately, the value $m_+ = 0.92$.
We do not investigate here what happens below this
value, where one enters a chaotic region with
the stationary measure dominated by several typical 
configurations. We will report results on the chaotic
region elsewhere. 

\medskip

\noindent
{\bf Acknowledgements.}
The authors wish to thank A. De Masi and E. Presutti
who inspired this work and supported our research with illuminating discussions.
M.C. acknowledges useful discussions with 
M. Kr\"{o}ger
on the implementation of Monte Carlo simulations.
We acknowledge financial supports
from Fondo di Ateneo per la Ricerca 2015 and 2016 (UniMoRe).
Part of this work was done during the authors stay at the Institute Henri Poincar\'e during the trimester  
``Stochastic Dynamics Out  of Equilibrium''.

\end{document}